\documentclass[12pt]{iopart}
\usepackage{iopams}  
\usepackage{graphicx}

\begin{document}
\title{The Jacobi map for gravitational lensing: the role of the exponential map}
\author{Paulo H. F. Reimberg and L. Raul Abramo}
\address{Instituto de F\'isica, Universidade de S\~ao Paulo, CP 66318, 05314-970, S\~ao Paulo, Brazil}
\ead{reimberg@fma.usp.br}

\begin{abstract}
We present a formal derivation of the key equations governing gravitational 
lensing in arbitrary space-times, starting from the basic properties of 
Jacobi fields and their expressions in terms of the exponential map.
A careful analysis of Jacobi fields and Jacobi classes near the origin of a 
light beam determines the nature of the singular behavior of the optical 
deformation matrix. We also show that potential problems that could arise 
from this singularity do not invalidate the conclusions of the original 
argument presented by Seitz, Schneider \& Ehlers (1994).
\end{abstract}
\vspace{2pc}
\submitto{\CQG}

\section{Introduction}

Gravitational lensing can be described, starting from very basic principles, 
in terms of the deviations of null geodesics with respect to a ``fiducial'' ray in a 
light beam. The formal groundworks upon which this description is
based were first spelled out by Seitz, Schneider and Ehlers (1994) \cite{seitz}, 
and their derivation has been widely used ever since  
\cite{uzan_lens, lewis_challinor, scalar_tensor, bartelmann_2010,bartelmann_schneider}. 

The fundamental objects in this description are the separation vectors 
$\boldsymbol{\xi}$, which determine how a beam of geodesics starting (or ending)
at a given point deviates from the fiducial. 
The relevant components of these vectors naturally belong to a 
2-dimensional space-like 
surface (the \emph{screen}) which is orthogonal to the direction of propagation of 
the null fiducial geodesic. And since, by construction, the beam is focused on 
the reference point, the separation vectors are such that 
$\boldsymbol{\xi} (0)=\boldsymbol{0}$.
This description is time-symmetric, in the sense that the reference 
event can be regarded either as the original source of the beam
(in which case the affine parameter is future-oriented), or as an observation 
event (in which case the affine parameter is past-oriented).

The separation vectors are in fact the projection on the screen of  
Jacobi fields along the fiducial ray.
A fundamental result in General Relativity is the fact that, to linear order in small 
perturbations around the fiducial geodesic, the Jacobi equation is \emph{linear}.
When projected on the screen, that equation leads to 
$\boldsymbol{{\xi}''} (\lambda) = {\mathcal{T}} (\lambda) \, 
\boldsymbol{\xi} (\lambda)$ , where
${\mathcal{T}}$ is called the \emph{optical tidal matrix}, and primes denote 
derivatives with respect to the affine parameter $\lambda$ along the fiducial geodesic. 
The separation of a null geodesic from the fiducial ray at any given value 
of the affine parameter, $\boldsymbol{\xi}(\lambda)$, would then be given by the action 
of a linear map (the \emph{Jacobi map},  \cite{seitz}) 
on the velocity of separation of that 
geodesic at the reference point, $\boldsymbol{{\xi}'} (0) =: \boldsymbol{\theta}$.
Ultimately, these two facts together allow us to frame gravitational lensing entirely 
in terms of the deviations $\boldsymbol{\theta}$ on the screen at the reference point.

We point out that previous demonstrations of these fundamental results have 
relied on a flawed argument which, if taken at face value, would imply
that the projections of the Jacobi fields on the screen would vanish identically. 
E.g., the argument presented in  \cite{seitz} is the following: since the Jacobi 
equation is linear, the projection on the screen of the Jacobi fields,
$\boldsymbol{\xi}(\lambda)$, can be related to the initial deviation
$\boldsymbol{\xi}'(0) = \boldsymbol{\theta}$ by a  linear transformation,
$\boldsymbol{\xi}(\lambda) = \mathcal{D} (\lambda) \boldsymbol{\theta}$. 
Substituting this expression back in the Jacobi equation, they are then 
able to derive the second-order linear differential equation which governs 
the evolution of the linear operator $\mathcal{D}(\lambda)$. 

They also claim that the linear operator $\mathcal{D}$ obeys the 
first-order differential equation
${\mathcal{D}}' = \mathcal{S} \,  \mathcal{D}$, where $\mathcal{S}$ is called \emph{optical deformation matrix} and is given in terms of optical observables for beams of null geodesics. This would imply, however, because the deviation can be
written as $\boldsymbol{{\xi}'} = {\mathcal{D}}' \, \boldsymbol{\theta}$, that 
$\boldsymbol{{\xi}'} = \mathcal{S} \, \boldsymbol{\xi}$ for all
values of the affine parameter. The problem with this last identity is, of course,
that a geodesic with $\boldsymbol{\xi}(0)=\boldsymbol{0}$ would
necessarily also have to satisfy $\boldsymbol{{\xi}'}(0)=\boldsymbol{0}$,
which would then imply that $\boldsymbol{\xi}=\boldsymbol{0}$ for all values of
the affine parameter. The only possible fix for this construction is for the 
optical deformation matrix to be {\emph{singular}} at the reference point, 
in precisely the right way to introduce some geodesic deviation 
$\boldsymbol{{\xi}'}$ at that point -- but this then constitutes an incomplete 
argument.

Here we will close the loophole in this argument. First, we will clarify the statement that the linearity of the Jacobi equation leads to the Jacobi map. Although our argument also relies on the linearity of the Jacobi equation, we present an explicit construction of the relation $\boldsymbol{\xi}(\lambda) = \mathcal{D}(\lambda) \boldsymbol{\theta}$ where 
$\mathcal{D}(\lambda)$ is directly determined from the \emph{exponential map}. 
Second, by employing the notion of Jacobi tensors we will show how to obtain a second-order linear differential equation for $\mathcal{D}$ which depends only on the Ricci and Weyl 
curvature tensors.

In our demonstration we will use a few basic results from Lorentzian geometry, in particular the fact that basis vectors along null geodesics cannot be orthonormal, and decomposition of vectors and tensor in terms of these bases are not defined in the classical sense. We follow \cite{hawking_ellis} in our construction of such a basis, over which separation vectors can be expressed. For this basis, we need to introduce the notions of the quotient by an equivalence relation. This then allows us to employ, in Section \ref{jacobi_classes_sec}, Jacobi classes and Jacobi tensors \cite{beem_ehrlich_easley}, which turn out to be key to our argument.

In Section \ref{jacobi_map_sec} we construct the Jacobi map, and obtain the differential equation that is satisfied by the operator $\mathcal{D}(\lambda)$. After writing the basic equations that 
govern the lensing problem in terms of the exponential map, we reexamine the argument presented by \cite{seitz}. Finally, we present some comments about conjugate points and critical behavior of beams of null geodesics.

Our conventions are as follows. Space-time is assumed to be a 
Lorentzian manifold $(M, g)$, and the tangent
space to a point $p \in M$ will be denoted by $T_pM$. 
Our metric has signature $-, +, +, +$, and contractions with it are denoted as 
$\langle . , . \rangle$, i.e., $\langle x, y \rangle = g_{\alpha \beta} x^{\alpha} y^{\beta}$. 
Greek indices range from $0$ to $3$ and latin indices from $1$ to $3$, 
except the indices $i$ and $j$, which can only take the values $1$ or $2$. 
The $2 \times 2$ identity matrix will be denoted by $\boldsymbol{1}$. 
The Riemann and Ricci tensors are denoted by $R(\, . \, , \, . \, ) \, . \,$
and $\mathrm{Ric}(\, . \, , \, . \, )$, respectively.

\section{The exponential map and Jacobi fields}

\begin{figure}
\begin{center}
\includegraphics[scale=0.5]{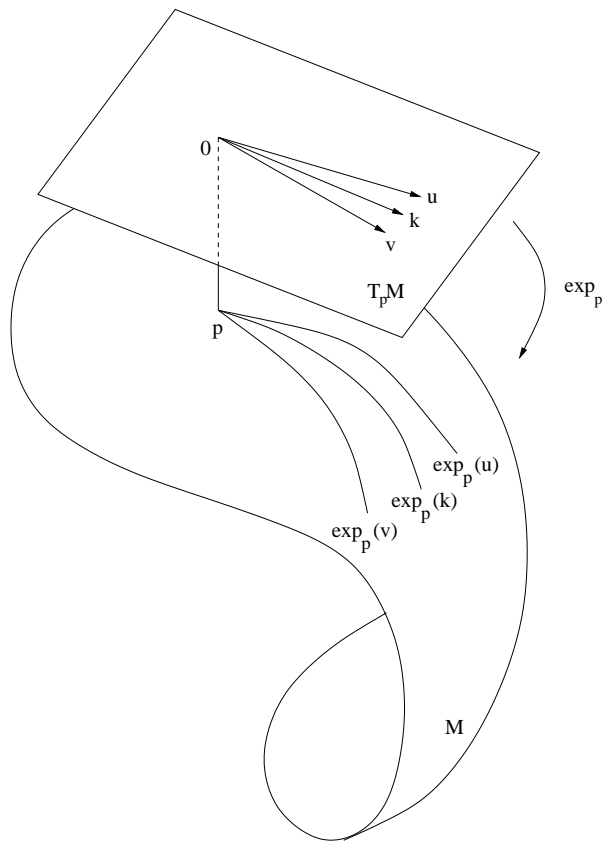}
\caption{\footnotesize Action of the exponential map. Given a vector in $T_pM$, its image under the exponential map is the point in $M$ obtained after a displacement of length $1$ along the (unique) geodesic starting at $p$ with a that vector as velocity. We show the image of the point 
$p$ under the exponential map for three different initial velocities  $u, k, v \in T_pM$.}
\label{exp_map}
\end{center}
\end{figure}

Let $\gamma: [0, 1] \to M$ be a null geodesic, so, by definition, 
$\frac{D}{d \lambda} \left( \frac{d \gamma}{d \lambda} \right) = 0$ and 
$\left \langle \frac{d \gamma}{d \lambda}, \frac{d \gamma}{d \lambda} \right \rangle = 0$. 
For given initial conditions $\gamma(0)=p$, and $\gamma'(0) = k \in T_pM$, the solution of the geodesic equation leads to a curve $\gamma(\lambda, p, k)$. If the solution $\gamma$ is defined, the affine parameter and the initial velocity can be scaled in such a way that $\gamma(1, p, k) \in M$. The point $\gamma(1, p, k) \in M$ is called the image of $k$ under the exponential map, for $p$ fixed. In our notation, the \emph{exponential map} $\mathrm{exp}_p : T_pM \to M$ is defined for a 
given $k \in T_pM$ as the point obtained along the unique geodesic in $M$ with $\gamma(0, p, k)=p$ and $\gamma'(0, p, k)=k$, after displacement by a length of $1$ along this curve, i. e., 
$\mathrm{exp}_p \, (k) := \gamma(1, p, k)$ \cite{beem_ehrlich_easley}. An illustration of the action of the exponential map is shown in figure \ref{exp_map}. 

Determining the explicit form of the exponential map is, in general, a very difficult task, and relies on the knowledge of all geodesics for a given space-time. Even without explicit formulas for the exponential map, however, much can be learned from its general properties. The geodesic homogeneity lemma, for example, establishes the aforementioned scaling relation between the affine parameter and the velocity, i. e., $\gamma(1,p,k)=\gamma(1/\lambda,p,\lambda k)$, and allows us to calculate the derivative of the exponential map at $\lambda=0$:
for any $k \in T_pM$,
\begin{equation}
\label{diffeo}
\frac{d}{d \lambda} [\mathrm{exp}_p(\lambda k)]_{\lambda=0} =
\frac{d}{d \lambda} [\gamma(1, p, \lambda k)]_{\lambda=0} = 
\frac{d}{d \lambda} [\gamma(\lambda, p, k)]_{\lambda=0} = k \, .
\end{equation}

The link between exponential map and variations of curves in the manifold is largely employed in the context of variational calculus \cite{perlick_fermat_i}, and can be established as follows: 
let $k(s)$, $s \in (-\varepsilon, \varepsilon) \subset \mathbb{R}$, be a curve in $T_pM$ such that $k(0)=k, \, dk/ds(0):=w$. Then $f(\lambda, s) := \gamma(1, p, \lambda k(s)) = \mathrm{exp}_p \, [\lambda k(s)]$ defines a parametrized surface in $M$. For $s=0$ we obtain the curve $\gamma(\lambda, p, k)$, and other values of the parameter $s$ generate variations of this original curve. A variational vector field $J$ along $\gamma$ constructed as $J(\lambda) = \frac{\partial f}{\partial s}(\lambda, s=0)$ is such that $J(0) = 0$ and $ J'(0) := \frac{\partial^ 2 f}{\partial \lambda \partial s}(\lambda, s=0) = w$. Such a vector field is known as \emph{Jacobi field} and satisfies the \emph{Jacobi equation} \cite{hawking_ellis}:
\begin{equation}
\label{Jacobi}
\frac{D^2 J}{d\lambda^2}+R(\gamma'(\lambda), J(\lambda))\gamma'(\lambda)=0 \; .
\end{equation}

Despite being broadly interpreted as geodesic deviation equation, the Jacobi 
equation has other solutions, depending on the initial conditions. For instance: if
$\gamma(\lambda)$ is a geodesic, then $\gamma'(\lambda)$ is a Jacobi field along $\gamma$, because $\gamma'(\lambda)$ is parallel-transported along itself, and because the Riemann curvature is anti-symmetric with respect to its two first arguments. Another example of a non-trivial Jacobi field along the curve $\gamma(\lambda)$ is $\lambda \gamma'(\lambda)$. 

\begin{figure}
\begin{center}
\includegraphics[scale=0.5]{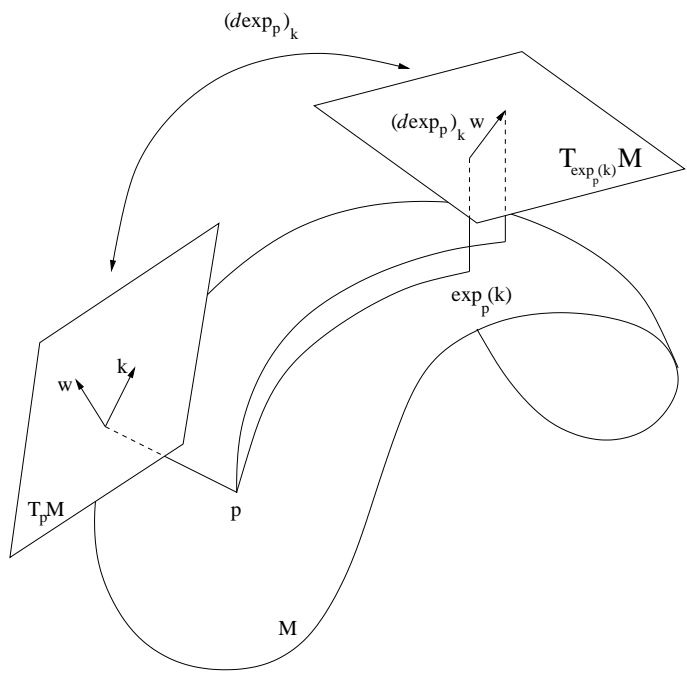}
\caption{\footnotesize The action of $(d \mathrm{exp}_p)_k$. 
For a given $w \in T_pM$, the action of $(d \mathrm{exp}_p)_k$ gives a 
vector in the tangent space of the point $\mathrm{exp}_p(k) \in M$. 
The image of $w$ under $(d \mathrm{exp}_p)_k$ is interpreted as 
the separation, at $\lambda =1$, of a fiducial null geodesic from 
another null geodesic starting at the same point but with an initial 
velocity of separation $w$. Here we identify $T_k(T_pM)$ with $T_pM$.}
\label{dexp}
\end{center}
\end{figure}

Describing gravitational lensing, however, requires to take into account deformations 
of beams of null geodesics that begin at a given specific point in space-time and, therefore,
conduce us to consider Jacobi fields that obey initial conditions $J(0)=0$ and $J'(0)=w$, 
where $w$ is the initial velocity of separation between a given geodesic in the 
beam and the fiducial one.  A fundamental property of these Jacobi fields, 
which lies at the heart of our discussion, is that the unique Jacobi field along 
$\gamma$ with $J(0)=0$ and $J'(0)=w$ is given by  -- see, e.g., 
Prop. 6 in chapter 8 of  \cite{oneill}, or Prop 10.16 of \cite{beem_ehrlich_easley}:
\begin{equation}
\label{jacobi_exp}
J(\lambda)= \lambda (d \mathrm{exp}_p) _{\lambda k} \,  w   \, .
\end{equation}
To make more clear the notation employed in   \ref{jacobi_exp} and show that it 
defines, in fact, a Jacobi field, we can explicitly calculate:
\begin{eqnarray}
J(\lambda) & = & \frac{\partial f}{\partial s}(\lambda, s) \vert_{s=0} = \frac{\partial}{\partial s} \mathrm{exp}_p(\lambda k(s))|_{s=0}
\nonumber \\ & = & (d\mathrm{exp}_p)_{\lambda k(0)} \left( \lambda \frac{dk}{ds}(s=0) \right) = \lambda (d\mathrm{exp}_p)_{\lambda k} w \, ,
\end{eqnarray}
where we have used the chain rule to establish the third equality. Expressing the exponential map in terms of its definition, we can write $\lambda (d\mathrm{exp}_p)_{\lambda k} w = \frac{\partial}{\partial s} \gamma(1, p, \lambda k(s)) \vert_{s=0}$. The differential of the exponential map $(d \mathrm{exp}_p)_{\lambda k}$ is the linear map 
\begin{displaymath}
(d \mathrm{exp}_p)_{\lambda k} : T_{\lambda k}(T_pM) \to T_{\mathrm{exp}_p(\lambda k)}M \, 
\end{displaymath}
whose matrix form is given by the directional derivatives of $\mathrm{exp}_p$ in the direction of the basis vectors of $T_{\lambda k}(T_pM)$, evaluated at $\lambda k$. In general, for $v \in T_pM$, the tangent space $T_v(T_pM)$ is defined as:
\begin{displaymath}
T_v(T_pM) := \{ \Phi_w : \mathbb{R} \to T_pM \} \, , \qquad \Phi_w(t) = v + sw \, \qquad v, w \in T_pM \, .
\end{displaymath}
Therefore, the space $T_v(T_pM)$ is isomorphic to $T_pM$, so for our purposes the two can be identified. Consequently, $(d \mathrm{exp}_p)_{\lambda k}$ will be understood as a linear map between the spaces $T_pM$ and $T_{\mathrm{exp}_p(\lambda k)}M$. In figure \ref{dexp} we illustrate the action of the differential of the exponential map on vectors in the tangent space of a point $p \in \gamma$.

To determine the behavior of the differential of the exponential map near the origin,
we shall rewrite   \ref{diffeo}, inserting an intermediate equality (application of the chain rule) between the first and the last one:
\begin{displaymath}
\frac{d}{d \lambda} [\mathrm{exp}_p(\lambda k)]_{\lambda=0} = (d\mathrm{exp}_p)_0 \, k =k \, .
\end{displaymath}
In other words, the differential of the exponential map calculated at the origin equals the identity matrix. This can be rephrased as the statement that the exponential map defines a diffeomorphism of a neighborhood of the origin of $T_pM$ into an open subset of $M$. This property also allows us to verify that $J$ given in   \ref{jacobi_exp} does satisfy $J'(0)=w$:
\begin{equation}
\label{J'w}
\fl J'(0) = \frac{D}{d\lambda} [\lambda(d \mathrm{exp}_p) _{\lambda k} \,  w]_{\lambda=0} = (d \mathrm{exp}_p) _{\lambda k} \,  w |_{\lambda=0}  +
\lambda \frac{D}{d \lambda} [(d \mathrm{exp}_p) _{\lambda k} \,  w ]_{\lambda=0} =w \, .
\end{equation}

\section{Construction of screen vectors}

We now turn our attention to the Jacobi fields that can be expressed as in 
  \ref{jacobi_exp}, and the way they can be decomposed in terms of 
a vectors basis along the null geodesic $\gamma$. 
Let's consider the vector basis along a null geodesic as introduced in Sec. 4.2 of 
\cite{hawking_ellis}: since $p=\gamma(0)$, we define a non-orthonormal basis 
$\{E_0, E_1, E_2, E_3 \}$ at $T_pM$ by requiring that $E_0 = \gamma'(0)$. 
By construction, $\langle E_0, E_0 \rangle = 0$, and we take $E_3$ to be
another null vector that obeys $\langle E_0, E_3 \rangle = -1$. The remaining 
vectors $E_1$, $E_2$ are then space-like and can be normalized. To extend 
this basis along $\gamma$ 
we simply parallel-transport the basis vectors.

With the help of this basis along $\gamma$, we can describe the separation 
of a bundle of geodesics that emerge from $p$ with any velocity of separation. 
We will be primarily concerned with the deformation of light beams with respect to
the fiducial ray $\gamma$, and will describe these other geodesics in terms of the
basis defined in the tangent space of all points along the fiducial curve. 
Since the separation of geodesics is described by the Jacobi fields, 
and all Jacobi fields that satisfy $J(0)=0$, $J'(0)=w$ have the form given 
in   \ref{jacobi_exp}, then
all the dependence on the separation between the geodesics is codified in $J'(0)$. 
If we decompose $J'(0)$ in the basis $\{ E_0, E_1, E_2, E_3 \}$,
the component $J_3'(0)$ is such that $\langle J_3'(0), \gamma'(0) \rangle \neq 0$. 
Hence, it follows from Gauss' Lemma that 
$\langle J_3 (\lambda), \gamma'(\lambda) \rangle \neq 0$ for all $\lambda$, 
where $J_3$ is the component of $J$ in the direction $E_3$. 
This component brings no information whatsoever about the problem 
we are interested in, so we shall only consider the vector space that results from 
the inverse image of $0$ by the application $\langle \, . \, , \gamma'(\lambda) \rangle$. 
With a notational abuse, we shall call this space $\tilde{T}_pM$, even if $p$ 
was previously assumed to be fixed.

\begin{figure}
\begin{center}
\includegraphics[scale=0.5]{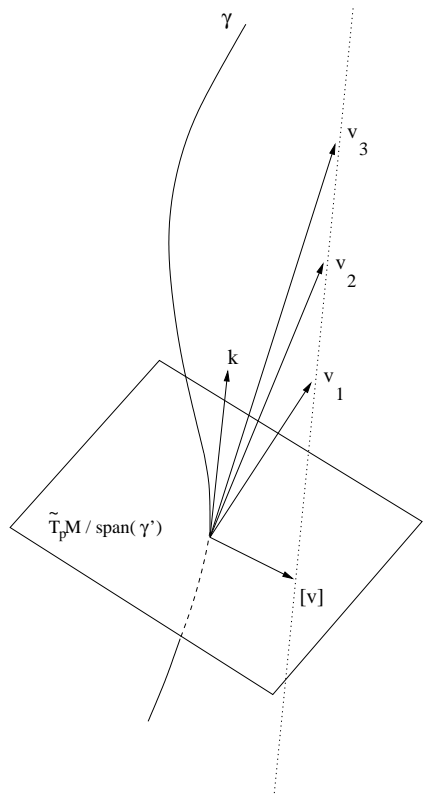}
\caption{\footnotesize Intuitive idea behind the construction of the screen. 
Any vector to which a scalar multiple of $k$ is added, will represent the same 
spatial separation from $k$. They should therefore be not distinguished for 
the sake of our problem and, for that reason, they are all identified inside a 
class of equivalence. The set of linearly independent equivalence classes 
forms $\tilde{T}_pM/\mathrm{span}(\gamma')$.}
\label{proj}
\end{center}
\end{figure}

There is, however, still one null Jacobi field remaining in $\tilde{T}_pM$: as we have already remarked, $\gamma'(\lambda)$ is itself a Jacobi field along $\gamma$, but it does not describe any spatial separation between geodesics. In order to eliminate this component, consider two vectors $W$ and $V$ in $\tilde{T}_pM$, and define $W \thicksim V$ if $W-V \in \mathrm{span}
(\gamma')$. It is easy to show that this is an equivalence relation in $\tilde{T}_pM$. The set of equivalence classes in $\tilde{T}_pM$ by the relation $\thicksim$ inherit the structure of vector space from $\tilde{T}_pM$. We shall denote $\tilde{T}_pM/\mathrm{span}(\gamma')$ the set of equivalence classes of $\tilde{T}_pM$ by the relation $\thicksim$. Given some $V \in \tilde{T}_pM$, $[V] \in \tilde{T}_pM/\mathrm{span}(\gamma')$ will denote the class corresponding to it. The same vector $[V]$ corresponds to the family of vectors $V + \alpha \gamma' \in \tilde{T}_pM$, $\alpha \in \mathbb{R}$, as illustrated in figure \ref{proj}. The two-dimensional space $\tilde{T}_pM/\mathrm{span}(\gamma')$ we have just constructed is called  \emph{screen}, and its elements are called \emph{screen vectors}. 

The result of the construction above is a two-dimensional vector space 
associated to every point along $\gamma$, in terms of which separations 
of geodesics in the beam can be described. All vectors in this space are 
space-like, and correspond to the naive idea of projection to a space 
``orthogonal'' to $\gamma'(\lambda)$ -- after taking into account that 
$\gamma'(\lambda)$ is orthogonal to itself.

\section{Jacobi classes, Jacobi tensors and optical scalars}
\label{jacobi_classes_sec}

We have just constructed a vector space to which the objects we want 
to describe will be restricted. We should ask how the space-time geometry 
is expressed when restricted to this space. Firstly, we note that, given 
$U, V \in \tilde{T}_pM$, then, since $\tilde{T}_pM$ has the structure of 
$\tilde{T}_pM/\mathrm{span}(\gamma') \oplus \mathrm{span}(\gamma')$,  
$U=[U]+\alpha k$ and $V=[V] + \beta k$ for same $\alpha, \beta \in \mathbb{R}$.
Therefore $\langle U, V \rangle = \langle [U] + \alpha k, [V] + \beta k \rangle = \langle [U], [V] \rangle$.  
Similarly, $\triangledown_{\gamma'} [U] = [\triangledown_{\gamma'} U]$. 
Also $[R(U, \gamma') \gamma'] = [R([U], \gamma') \gamma' + \alpha R(\gamma', \gamma') \gamma'] = [R([U], \gamma') \gamma'] :=[R]([U], \gamma') \gamma'$, due to the symmetries of Riemann tensor. 

We shall now consider the Jacobi fields restricted to the quotient space or, \emph{Jacobi classes}, as these restrictions are usually called. In general, Jacobi classes are smooth vector fields in the quotient space satisfying the relation:
\begin{equation}
\label{class}
j'' + [R] (j, \gamma') \gamma' = [\gamma'] \, .
\end{equation}
It can be shown that, given a Jacobi class $j \in \tilde{T}_pM/\mathrm{span}(\gamma')$, there exists a Jacobi field $J$ in $\tilde{T}_pM$ such that $[J]=j$, and, conversely, if $J$ is a Jacobi field in $\tilde{T}_pM$, $j=[J]$ is a Jacobi class in $\tilde{T}_pM/\mathrm{span}(\gamma')$ \cite{beem_ehrlich_easley}.

We now introduce objects called \emph{Jacobi tensors} \cite{beem_ehrlich_easley}, which are
of fundamental importance in the discussion about optical properties of light beams, and in terms of which the \emph{optical scalars} will be defined. Let $A$ be a bilinear form $A : \tilde{T}_pM/\mathrm{span}(\gamma') \to \tilde{T}_pM/\mathrm{span}(\gamma')$. For any $W \in \tilde{T}_pM$, $[R]A([W]) := [R](A[W], \gamma')\gamma'$. We shall say that $A$ is a Jacobi tensor if:
\begin{equation}
\label{jacobi_tensor}
A'' + [R] A = 0 \; ,
\end{equation}
subject to convenient initial conditions, and
\begin{equation}
\label{kernel}
\mathrm{ker}\{ A(\lambda) \} \cap \mathrm{ker} \{ A'(\lambda) \} 
= [\gamma'(\lambda)] \, .
\end{equation}

If the bilinear form $A$ satisfies   \ref{jacobi_tensor}, then vector fields constructed by its action on vector basis $E_i$ will be Jacobi classes. The condition  \ref{kernel} eliminates trivial Jacobi classes from consideration: if $j_i = A E_i$ are Jacobi classes along $\gamma$, since $E_i$ are parallel propagated, $j_i'=A' E_i$ give their (covariant) derivatives. If at some point along $\gamma$ the same linear combination of $E_i$s belong to the kernel of both $A$ and $A'$, then the action of $A$ on this linear combination will give rise to a trivial Jacobi class. To see that this is the case, we should recall that the dimension of the kernel of $A$ is only non-null at conjugate points \cite{beem_ehrlich_easley}, and conclude, by taking a point $\lambda^*$ conjugate to $\lambda=0$ along $\gamma$ for which $j'(\lambda^*)=[\gamma']$, that the only possible solution for   \ref{class} in this case does not represent separation of curves in a beam.

The condition  \ref{kernel} has an important consequence: objects defined by  
$A' \, A^{-1}$ may be singular either because $A$ is singular, or because 
$A'$ is singular, but never because both are singular at the same point. 

The optical scalars \emph{expansion}, \emph{vorticity} and 
\emph{shear} of a geodesic beam are defined in terms of the Jacobi tensors, 
namely, through combinations of the object $ B =: A' \, A^{-1}$. 
Explicitly, we have \cite{beem_ehrlich_easley}:

\begin{itemize}
\item[(i)] \emph{expansion} 
\begin{equation}
\label{expansion}
\Theta = \mathrm{Tr} (  B ) \; ,
\end{equation}

\item[(ii)] \emph{vorticity}
\begin{equation}
\label{vorticity}
\omega = {\rm Im} \, (B)  \; ,
\end{equation}

\item[(iii)] \emph{shear}
\begin{equation}
\label{shear}
\sigma = {\rm Re} \, (B)
- \frac{\Theta}{2} \boldsymbol{1} \; .
\end{equation}
\end{itemize}
In terms of these objects the Raychaudhuri equation reads:
\begin{equation}
\Theta' = - \mathrm{Ric}(\gamma', \gamma') - \mathrm{Tr}(\omega^2)
- \mathrm{Tr}(\sigma^2) - \frac{\Theta^2}{2} \, .
\end{equation}

Jacobi tensors are, then, key objects to understand and describe deformations 
of light beams. Next we will address the explicit construction of a Jacobi tensor 
that satisfies the physical restriction: since we are interested in gravitational 
lensing, all geodesics in the beam start at the same point and have different
separation velocities with respect to a fiducial ray.

\section{The Jacobi map}
\label{jacobi_map_sec}

In what follows we will explicitly build the Jacobi map. The argument of Ref. \cite{seitz} is that the linearity of the Jacobi equation implies that the Jacobi map should hold -- although this is in fact true, we will see that it is far from trivial.

By definition, in $\tilde{T}_pM/\mathrm{span}(\gamma')$,
\begin{displaymath}
j_i(\lambda) = \langle [J(\lambda)], [E_i(\lambda)] \rangle 
= \langle J(\lambda), E_i(\lambda) \rangle 
= \langle (\lambda d \mathrm{exp}_p)_{\lambda k}   w , E_i(\lambda) \rangle \, ,
\end{displaymath}
where $w=J'(0)$ and $i=1, 2$. This initial velocity can be expressed as $w = \theta_1 E_1(0) + \theta_2 E_2(0) + \alpha E_0$ and, since $\langle (\lambda d \mathrm{exp}_p)_{\lambda k}   E_0, 
E_i(\lambda) \rangle = 0$, there is no loss of generality in taking $w$ to be restricted to 
$\tilde{T}_pM/\mathrm{span}(\gamma')$. This just means that we only have to determine the space-like components of the initial velocity of dispersion of geodesics in order to determine their future evolution. The tangential component of the geodesics may also deserve attention, 
for instance in cosmological contexts where redshifts or the Sachs-Wolfe \cite{sachs_wolfe} effect can take place, but in those cases the corrections can be treated separately.

Taking the initial spread velocity restricted to $\tilde{T}_pM/\mathrm{span}(\gamma')$, then $\langle [\lambda \, (d \mathrm{exp}_p)_{\lambda k} \, . \, ], [\, . \, ] \rangle$ constitutes a bilinear form in $\tilde{T}_pM/\mathrm{span}(\gamma')$, and therefore it admits a matrix representation:
\begin{equation}
\label{jacobi_matrix}
\fl \left( \begin{array}{c}
j_1 \\ j_2
\end{array} \right)=
\left( \begin{array}{cc}
\langle \lambda \, (d \mathrm{exp}_p)_{\lambda k}  E_1(0) , E_1(\lambda) \rangle & 
\langle \lambda \, (d \mathrm{exp}_p)_{\lambda k}  E_1(0) , E_2(\lambda) \rangle \\
\langle \lambda \, (d \mathrm{exp}_p)_{\lambda k}  E_2(0) , E_1(\lambda) \rangle & 
\langle \lambda \, (d \mathrm{exp}_p)_{\lambda k}  E_2(0) , E_2(\lambda) \rangle
\end{array}\right)
\left( \begin{array}{c}
\theta_1 \\\theta_2
\end{array} \right) \, . 
\end{equation} 

\ref{jacobi_matrix} expresses the separation of null geodesics in terms of
the initial velocity of dispersion of the beam. If we denote the $2 \times 2$ matrix 
of   \ref{jacobi_matrix} by $\mathcal{D}(\lambda)$, even if it depends on 
the point $p$ ($\lambda=0$) and on the fiducial geodesic $\gamma$, 
we obtain \footnote{ Ref. \cite{seitz} calls $\boldsymbol{\xi}$ what we are calling 
$\bold{j}$.}:
\begin{equation}
\label{jacobi_map}
\bold{j} = \mathcal{D}(\lambda) \, \boldsymbol{\theta} \, .
\end{equation}

This is, in fact, what is usually called \emph{Jacobi map} in the gravitational lensing
literature \cite{seitz}. Using the explicit form given in   \ref{jacobi_matrix}, it is easy 
to see, after   \ref{jacobi_exp},  \ref{diffeo} and  \ref{J'w}, that the matrix $\mathcal{D}$ satisfies the initial conditions $\mathcal{D}(0)=0$ and $\mathcal{D}'(0) = \boldsymbol{1}$.

\section{A differential equation for $\mathcal{D}$}

We will now show that the matrix $\mathcal{D}$ which appears in the Jacobi map,  \ref{jacobi_map}, satisfies the differential equation  \ref{jacobi_tensor}. This can be verified with the help of the Jacobi equation  \ref{Jacobi}.

Let's define $\epsilon = [E_1] + i[E_2]$, $\mathcal{J} = j_1 + i j_2$, and let $\bar{\epsilon}$ and $\bar{\mathcal{J}}$ be their complex conjugates. Since for a given Jacobi field there is a unique Jacobi class associated with it, 
\begin{eqnarray}
\label{J''}
\langle \epsilon, [J]'' \rangle & = & - \langle \epsilon, [R(J, \gamma')\gamma'] \rangle 
=  -\mathcal{J} \frac{1}{2} \langle \epsilon, [R] (\bar{\epsilon}, \gamma' )\gamma' \rangle \nonumber\\ & &
- \bar{\mathcal{J}} \frac{1}{2} \langle \epsilon, [R](\epsilon, \gamma') \gamma' \rangle  = - \mathcal{R} \mathcal{J} - \mathcal{F} \bar{\mathcal{J}} \, 
\end{eqnarray}
where $\mathcal{R}$ and $\mathcal{F}$ are given in terms of the Riemann curvature. Writing $k = \gamma'$, it can be shown \cite{straumann} that these objects are given by
\begin{equation}
\mathcal{R}=\frac{1}{2} \langle \epsilon, [R](\bar{\epsilon}, k)k \rangle=\frac{1}{2} R_{\mu \nu} k^{\mu} k^{\nu} \, ,
\end{equation}
and
\begin{equation}
\mathcal{F} = \frac{1}{2} \langle \epsilon, 
[R](\epsilon, k)k \rangle = \frac{1}{2} \epsilon^{\alpha} \epsilon^{\beta}  C_{\alpha \mu \beta \nu} k^{\mu} k^{\nu} \, ,
\end{equation}
where $R_{\mu \nu}$ and $C_{\alpha \mu \beta \nu}$ are the components of the Ricci and Weyl tensor, respectively. 

Substituting   \ref{jacobi_map} in   \ref{J''} we obtain, indeed,   \ref{jacobi_tensor}. Explicitly,
\begin{equation}
\label{D''}
\mathcal{D}''(\lambda) + \mathcal{T}(\lambda) \mathcal{D}(\lambda) = 0
\end{equation}
where $\mathcal{T}$ is the matrix form of the endomorphism $[R](\, . \, , \gamma')\gamma'$ acting on $\tilde{T}_pM/\mathrm{span}(\gamma')$, and is usually called \emph{optical tidal matrix}. In terms of $\mathcal{R}$ and $\mathcal{F}$, $\mathcal{T}$ reads:
\begin{equation}
\mathcal{T} = \left( \begin{array}{cc}
\mathcal{R} + \mathrm{Re}\mathcal{F} & \mathrm{Im} \mathcal{F} \\
\mathrm{Im} \mathcal{F} & \mathcal{R} - \mathrm{Re} \mathcal{F}
\end{array} \right) \, . 
\end{equation}

As already remarked,   \ref{D''} must be subjected to the initial conditions $\mathcal{D}(0)=0$ and $\mathcal{D}'(0) = \boldsymbol{1}$.

\section{The condition on the kernels}

In order to verify that $\mathcal{D}$ is in fact a Jacobi tensor, we must guarantee that the condition expressed by   \ref{kernel} is satisfied by the bilinear form constructed in   \ref{jacobi_matrix}, for all values of $\lambda$ for which $\gamma$ is defined. Since  
 \ref{kernel} holds for $\lambda=0$ because of the initial conditions, we shall consider the case $\lambda \neq 0$. Keeping in mind that the vectors $E_i(\lambda)$ are parallel-transported along the fiducial ray, taking the derivative $\mathcal{D}'(\lambda)$ leads to
\begin{eqnarray}
\label{D_prime}
\mathcal{D}'(\lambda) & = &
\left( \begin{array}{cc}
\langle  (d \mathrm{exp}_p)_{\lambda k} E_1(0) , E_1(\lambda) \rangle & \langle 
(d \mathrm{exp}_p)_{\lambda k} E_1(0) , E_2(\lambda) \rangle \\
\langle (d \mathrm{exp}_p)_{\lambda k} E_2(0) , E_1(\lambda) \rangle & \langle 
(d \mathrm{exp}_p)_{\lambda k} E_2(0) , E_2(\lambda) \rangle 
\end{array}\right) \nonumber\\ & + & 
\lambda \frac{d}{d \lambda}
\left( \begin{array}{cc}
\langle  (d \mathrm{exp}_p)_{\lambda k} E_1(0) , E_1(\lambda) \rangle & \langle 
(d \mathrm{exp}_p)_{\lambda k} E_1(0) , E_2(\lambda) \rangle \\
\langle  (d \mathrm{exp}_p)_{\lambda k} E_2(0) , E_1(\lambda) \rangle & \langle 
(d \mathrm{exp}_p)_{\lambda k} E_2(0) , E_2(\lambda) \rangle 
\end{array}\right) \, .
\end{eqnarray}

In order to determine if there are common elements in the kernels of $\mathcal{D}$ and $\mathcal{D}'$, we shall study if elements in the kernel of  $\mathcal{D}$ may also be in the kernel of $\mathcal{D}'$. Given its explicit form of   \ref{jacobi_matrix}, and the fact that 
Jacobi fields expressed by   \ref{jacobi_exp} are linearly independent if the initial velocities are linearly independent,  $\mathrm{ker}(\mathcal{D}(\lambda))$ has non-trivial elements only 
at points which are conjugate points to $\lambda=0$ along $\gamma$. 

\begin{figure}
\begin{center}
\includegraphics[scale=0.5]{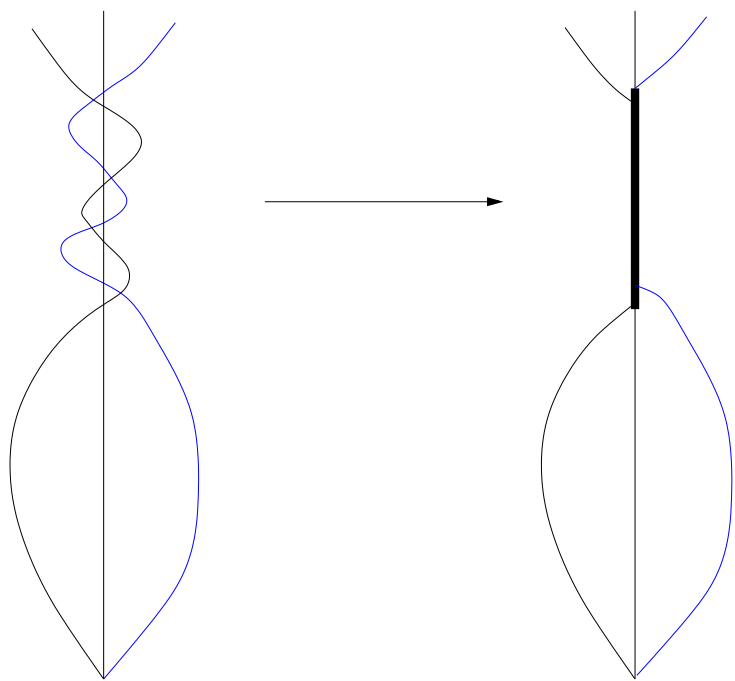}
\caption{\footnotesize Two independent Jacobi classes along a fiducial ray. 
Let's say that after a given value of the affine parameter along the fiducial 
ray, there is a sequence of conjugate points such that, between two 
conjugate points of the same Jacobi class, the second class passes 
through a point of maximum separation. A collapse of such a 
sequence of conjugate points in a neighborhood, depicted by the thick line 
in the diagram in the right, could give rise to a violation of condition  \ref{kernel}. This, however, is not possible because the set of conjugate points along $\gamma$ is finite, as states the Morse Index Theorem.}
\label{tranca}
\end{center}
\end{figure}

At conjugate points, the first term in the right hand side of   \ref{D_prime} is singular. The second term in the right hand side of   \ref{D_prime}, however, must not be singular for the 
same Jacobi class because, if it were, then the Jacobi class thus obtained would be the trivial one.
We must, therefore, make the derivative term in the right-hand side of   \ref{D_prime} to be singular because of the second Jacobi class, i.e., the second Jacobi class must reach an extremum exactly at the same value of the parameter where the first class vanishes. Although this is not sufficient to cause a violation of the condition expressed by   \ref{kernel}, if we could have a sequence of conjugate points along the two Jacobi classes, as illustrated in figure \ref{tranca} (i. e., between any two conjugate points of a class, there is a maximum of the other class), and if we were able to make all conjugate points accumulate in a neighborhood of some point in the interval, then there could be a way of taking this limit such that the condition  \ref{kernel} would fail in some point of that neighborhood. The Morse Index Theorem, however, states that the set of conjugate points along a null geodesic is a finite set \cite{beem_ehrlich_easley}. Consequently, an accumulation of conjugate points cannot take place, and hence the limit depicted in figure \ref{tranca} cannot exist. We conclude, then, that   \ref{kernel} holds for the bilinear form $\mathcal{D}$, and therefore it determines a Jacobi tensor. 

\section{The behavior near the origin and the sweep method}

We now return to the argument presented in \cite{seitz} to show that, 
despite the loophole in their argument leading to an equation equivalent 
to   \ref{D''}, this is of no consequence. Specifically, we will show 
that a matrix $\mathcal{S}$ such that $\mathcal{D}'=\mathcal{S} \mathcal{D}$ 
and $\bold{j}' = \mathcal{S} {j}$ can be defined,
but that it cannot be taken as a starting point for obtaining   \ref{D''}.

Suppose that there exists a matrix $\mathcal{S}$ such that 
$\mathcal{D}'(\lambda)=\mathcal{S}(\lambda) \mathcal{D}(\lambda)$, 
and that we restrict the domain of $\lambda$ in such a way that there 
are no conjugate points to $\lambda=0$ along $\gamma$ . Then, 
from   \ref{D''} it follows that $\mathcal{S}' + \mathcal{S}^2 = T$. 
This equation is known as \emph{matrix Riccati equation}, and 
corresponds to the usual association of Riccati's equation to 
second-order homogeneous differential equations.

The initial conditions for this matrix Riccati equation should be carefully 
considered. If both $\mathcal{D}(0)=0$, $\mathcal{D}'(0)=\boldsymbol{1}$ 
and $\mathcal{D}'(\lambda)= \mathcal{S}(\lambda) \mathcal{D}(\lambda)$ hold
simultaneously, then $\mathcal{S}(0)$ must not be limited. Consequently, one 
cannot guarantee the existence of solutions for this first-order linear differential 
equation if $\lambda$ is in the neighborhood of the origin ($\lambda=0$), and 
in that case one cannot assure a solution for the associated Riccati equation either.

\begin{figure}
\begin{center}
\includegraphics[scale=0.5]{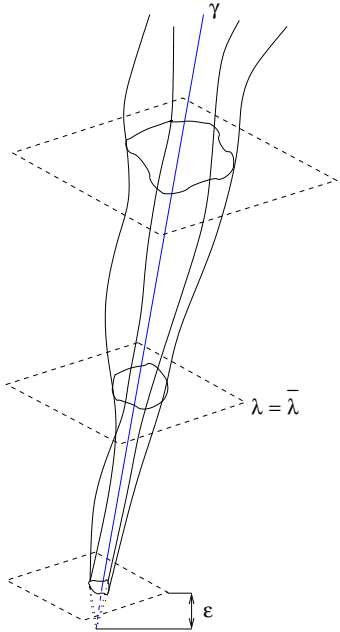}
\caption{\footnotesize Beam of null geodesics starting at a point. 
At $\lambda = \varepsilon$ the beam behaves like a congruence, 
and an equation like $\mathcal{D}'(\lambda)=\mathcal{S}(\lambda) \mathcal{D}
(\lambda)$ can be solved for given initial conditions, if $\mathcal{S}(\epsilon)$ is 
bounded. If we solve the same equation for initial conditions at 
$\lambda=\bar{\lambda}$, we can see if the solutions match and, if they do, 
then we can say that it is also a solution of the second-order equation 
$\mathcal{D}'' + T \mathcal{D}=0$.}
\label{beam}
\end{center}
\end{figure}

To avoid problems with the solution around $\lambda=0$, we take 
$\lambda > \epsilon > 0$ and impose initial conditions $\mathcal{D}(\varepsilon)$ 
and $\mathcal{D}'(\varepsilon)$, as illustrated in figure \ref{beam}. The 
equation $\mathcal{D}'=\mathcal{S} \mathcal{D}$ will then admit solutions 
for $\lambda > \varepsilon$, if $\mathcal{S}(\lambda)$ is bounded for 
$\lambda > \varepsilon$.
If some further conditions $\mathcal{D}(\bar{\lambda})$ and 
$\mathcal{D}'(\bar{\lambda})$, $\bar{\lambda} > \varepsilon$, are also 
given, then one can solve for $\mathcal{D}'=\mathcal{S} \mathcal{D}$, 
subjected to these conditions, and check whether the solutions match 
those obtained by imposing initial conditions at $\lambda=\varepsilon$. 
If the two solutions match, the initial conditions given at $\varepsilon$ and 
$\bar{\lambda}$ are said consistent, and the solution $\mathcal{D}$ will 
also be a solution of   \ref{D''} in the same range of parameters. 
This method of obtaining solutions is known as ``sweep method'' 
\cite{gelfand_variational}. 

The sweep method would work well if geodesics in the beam did not 
cross each other -- in other terms, if they formed a congruence.  If the 
beam emerges from the same point, however, initial conditions at that 
point cannot be used for the ``forward sweep'' step of the method, 
because one cannot guarantee the existence of solutions for the 
differential equation in this case. In spite of that, if one knows, by any 
other method, a solution for   \ref{D''}, then the behavior of this 
solution near $\lambda = 0$ can be investigated. We presented 
explicitly how $\mathcal{D}$ is given in terms of the exponential map 
in   \ref{jacobi_matrix}, and therefore we can investigate the 
properties of $\mathcal{D}(\varepsilon)$, $\varepsilon \approx 0$. 

If one recalls that the entries of   \ref{jacobi_matrix} are projections 
of Jacobi fields in base vectors, and that Jacobi fields in the 
neighborhood of the origin behave like $\varepsilon + {\cal{O}}\varepsilon^3$, 
we see that for $\varepsilon$ close to zero, 
$\mathcal{D}(\varepsilon) \approx \varepsilon \boldsymbol{1}$. 
Then, from   \ref{D_prime} we obtain that $\mathcal{D}'(\varepsilon) \approx \boldsymbol{1}$. 
Hence, an equation like $\mathcal{D}' = \mathcal{S} \mathcal{D}$ would only 
make sense if near $\lambda = 0$ the matrix $\mathcal{S}$ had a behavior 
like $\mathcal{S}(\varepsilon) \approx (1/\varepsilon) \boldsymbol{1}$.

We remark that, despite the fact that the equation  
$\mathcal{D}' = \mathcal{S} \mathcal{D}$ may not be inconsistent with the 
initial conditions $\mathcal{D}(0)=0$ and $\mathcal{D}'(0)=\boldsymbol{1}$, 
if $\mathcal{S}$ diverges in the prescribed way near the origin, the initial 
conditions for the Riccati equation cannot be provided at the origin, and 
the sweep forward cannot be employed. In other words, a general solution for 
  \ref{D''} must be known in order to verify that 
$\mathcal{D}' = \mathcal{S} \mathcal{D}$ is consistent near the origin -- but 
this last equation could never be used to derive or to generate solutions to 
  \ref{D''} if the origin (or conjugate points) are in the domain of the 
solution.

\section{Conjugate points and critical behavior}

The set of conjugate points along the geodesic $\gamma$ is not degenerate
when the restriction to the quotient space is taken: it is in fact completely 
preserved, because the Jacobi field in the direction of $\gamma'$ 
(i.e., $\gamma'$ itself) does not vanish.
All possible non-trivial Jacobi fields that satisfy $J(0)=0$, and vanish at some
other point $\gamma(\lambda_1)$, correspond to Jacobi classes
satisfying $[J(0)] = [\gamma'(0)]$ and $[J(\lambda_1)]=[\gamma'(\lambda_1)]$.
The conjugate points along $\gamma$ correspond to the critical points of the
matrix $\mathcal{D}$, and induce optical critical behavior which is codified by 
the diverging expansion given by   \ref{expansion}.

Because $\tilde{T}_pM/\mathrm{span}(\gamma')$ is bi-dimensional, the maximal
multiplicity of any conjugate point is two. After a conjugate point of multiplicity 
one, the source's image will appear inverted -- something that does not occur 
after conjugate points of multiplicity two (or two conjugate points of multiplicity 
one).

The presence of conjugate points along $\gamma$ is much more significant in aspects other than classifying an image's orientation.  \cite{perlick_criteria} showed that conjugate
points are necessary to form multiple images, and due to the connections between the existence of conjugate points along geodesics and the energy conditions \cite{hawking_ellis}, 
multiple images will always occur in a large class of space-times, as long as geodesics can be enough extended. As conjugate points are critical points of the energy functional, they 
also play an important role in the formulation of Morse Theory applied to light rays, in particular in the formulation of theorems on the possible number of images in lensing configurations 
\cite{perlick_geometric_view}.

\section{Discussion}

The description of gravitational lensing presented here is central in many applications, especially in cosmology  (\cite{seitz, uzan_lens, lewis_challinor, scalar_tensor}), when   \ref{D''} can be solved perturbatively to generate a lens map. Equations with the same general structure are also fundamental in the discussion of singularities in the Penrose limit \cite{blau} and can have other applications. Writing these equations in a rigorous way has allowed us to determine their explicit form in terms of the exponential map, beyond generating a clear interpretation for the Jacobi map and validate its centrality in the description of gravitational lensing.

The limitations of the derivation presented in this work should also be clarified. First, the Jacobi equation itself is only approximative in first order to the problem of geodesic spread. Corrections may be included, such as non-linearities with respect to the velocity dispersion \cite{perlick_gen_jacobi}. Second, despite these attempts to enlarge the limits of application of generalizations of the Jacobi equation, there are obstacles that may not be overpassed in this formulation, such as configurations in which \emph{cut points} \cite{beem_ehrlich_easley} exist, and a more careful geometrical analysis is required. One such example is lensing produced by a string, as introduced by \cite{vilenkin_lenses}, which induces a flat space-time that nevertheless generates multiple imaging. In full generality, gravitational lensing phenomena involve conjugate 
points and even cut points, and we can expect that an observer's past light cone will eventually reveal this complex geometry.
\ack
This work was supported by FAPESP.

\section*{References}
\bibliographystyle{unsrt}
\bibliography{referencias}

\end{document}